# Raman scattering, X-ray photoemission spectra and superconductivity of a tiny Ag diffusion to $MgCNi_3$

Fa–Min Liu[1,2], J. Q. Li[1], C. Dong[1], T. M. Wang[2], Y. Q. Zhou[1], H. Chen[1]

[1] National Laboratory for Superconductivity, Institute of Physics, Chinese Academy of Sciences, Beijing 100080, P. R. China

< Tel. 86-10-82649180, Fax. 86-10-82649531, E-mail: fmliu@aphy.iphy.ac.cn ; fmliu@etang.com >

[2] Center of Material Physics and Chemistry, Beijing University of Aeronautics & Astronautics, Beijing 100083, P. R. China

**Abstract**: The tiny Ag diffusion to $MgCNi_3$ has been prepared by solid states reaction. Its structure was characterized by X-ray diffractometer (XRD). The results show that a small mount of Ag substitute for Ni sites, and much of Ag are in vacancy sites of the $MgCNi_3$. We have further studied the surface properties of Ag-$MgCNi_3$ using Raman scattering spectra, and X-ray photoemission spectra (XPS). Raman spectroscopy shows that the Ag-$MgCNi_3$ has a special Raman peak around 842.1 $cm^{-1}$ compared to that of C. The superconductor transition temperature (about 6.6 K) of Ag-$MgCNi_3$ was lower than that of pure $MgCNi_3$. It was interpreted properly by the conventional BCS phonon mechanism.

1. Introduction

The discovery of superconductivity in the intermetallic antiperovskite structure $MgCNi_3$ [1] with the transition temperature of Tc–8.5K has stimulated scientist interest in perovskite structure and other unusual superconductivity phase such as the Ni, Pd and Pt borocarbides and boronitrides [2,3]. Much work has been done for $MgCNi_3$ in theory [4-6] and experiment [7,8]. Recently Hayward, et al. [9] reported that Ni-site doping in $MgCNi_3$ decreased the superconducting transition temperature Tc. They found that with Cu doping (electron doping) on the Ni site, Tc decreases systematically, but with Co doping (hole doping), the superconductivity disappears abruptly for doping of only 1%. The electronic structures of electron (Cu) and hole (Co) doping in $MgCNi_3$ [10] have been calculated using the self-consistent tight-binding linear muffin-tin orbital method. Their results show that electron (Cu) and hole (Co) doping of $MgCNi_3$ reconstructs its band structure but does not lead to magnetic order. Li, et al. [11] have recently reported that structural investigations on a series of superconducting materials with nominal compositions of $Mg_xC_yNi_3$ ($1.0 < x < 1.3$, $1.0 < y < 1.55$) have revealed evident structural inhomgeneity. They found that the presence of the local C deficiency could be a dominant factor affecting the crystal structure and superconductivity. Here we report that the Ag diffusion to $MgCNi_3$ reduced the superconductivity near 6.6 K and studied their surface properties using Raman scattering and X-ray photoemission spectra.

2. Experiment

Polycrystalline samples of Ag-$MgCNi_3$ were prepared by a conventional solid states reaction. The Ag-$MgCNi_3$ sample used in the measurements was prepared from Mg (99.5%) powder of 50 meshes, C (99.99%) powder of 325 meshes and Ni (99.8 %) powder of 325 meshes. The initial stoichiometric $MgCNi_3$ mixture of Mg, C and Ni was well mixed and pressed into pellets, then these pellets and high pure thin piece of Ag (99.9999 %) were wrapped in a Ta foil and sealed in a quartz ampoule. The samples were sintered at 900 $^oC$ for 2 hours. Then it was cooled down to room temperature. The phase structures were analyzed by an X-ray diffractometer (XRD) (MAC Science, M18X) using Cu-K$\alpha$ radiation. Transmission electron microscopy (TEM) studies were used to determine the microstructure of the samples with H-9000NA with an atomic resolution of about 0.19 nm. Specimens for transmission-electron microscopy (TEM) observations were polished mechanically with a Gatan polisher to a thickness of around 50$\mu$m and then ion-milled by a Gatan-691 PIPS ion miller for 3 h. Electrical resistance and the superconducting transition were investigated using the standard four probe method.

X-ray photoemission spectra data were obtained with a VG ESCALAB MK II spectrometer using Mg $K_\alpha$ source radiation. The working pressure in the XPS chamber was approximately $6 \times 10^{-7}$ Pa. Survey spectra were collected with pass energy of 50 eV. Spectra are measured at room temperature with photoemission 75$^o$ from the surface normal for the polycrystalline Ag-$MgCNi_3$. Instrument was calibrated with the Au 4f line. Specimens for XPS observations were ground on the abrasive paper (CW2000), and then were finished the surface on the finishing machinery.

3. Results and discussion

It is well known that the $MgCNi_3$ is a strong-coupling superconductor, and has the perovskite structure over the whole temperature range, with the lattice parameters a =

0.381221 nm at 295 K. The results of TEM observations will present in another paper. Fig. 1 shows the XRD of Ag-$MgCNi_3$. It shows that the cubic Ag contains in the cubic antiperovskite structure $MgCNi_3$. The lattice parameter of Ag was increased 4% and that of $MgCNi_3$ was decreased 1% estimated by XRD. This demonstrates that a small amount of Ag substrate for Ni sites, and the other Ag are in vacancy site maintained cubic phase Ag. We estimated that about 7.7% diffused to $MgCNi_3$ from XRD.

Fig. 2 shows Raman scattering of Ag-$MgCNi_3$. Three peaks appear at the position of 842.1, 1351.6 and 1587.9 $cm^{-1}$. Compared with Raman scattering of C, we know that the feature Raman peak for Ag-$MgCNi_3$ is at 842.1 $cm^{-1}$, which attributed to Mg-Ni-C bonds.

Fig. 3 shows Mg 2p core level of Ag-$MgCNi_3$. The core level of Mg 2p is at around 49.9 to 50.3 eV. This demonstrates that the metal magnesium and magnesium oxide coexist in the surface of Ag-$MgCNi_3$.

Fig. 4 shows C1s core level of Ag-$MgCNi_3$. One can see that the core level of C 1s is at around 284.6 eV, which is agreeable with Ref. [12]. In addition, we note that the C 1s line has a tail on the high binding energy side, which indicated that the C 1s electrons are strongly hybridized with Ni and Mg. Without C located at the center of cubic cell, $MgNi_3$ is a simple ordered intermetallic compound with the face-centered –cubic structure ($Cu_3Au$-type). With C, the Ni 3d band in $MgNi_3$ in MgNi3 is very narrow, leading to a magnetic ground state with Ni magnetic moment of 0.43 $\mu_B$ [6].

Fig. 5 indicates Ni 2p core level of Ag-$MgCNi_3$. We found that the core levels of Ni $2p_{1/2}$ and Ni $2p_{3/2}$ are at around 861.5 and 855.5 eV respectively, with a peak separation of 6.0 eV between those two peaks. Compared with the standard of nickel [12], we know that the nickel in the surface of Ag-$MgCNi_3$ is not metal state. In addition, the long tail on the low binding energy side of the Ni 2p core levels shows that the strong Ni-C covalent interaction [4], and the Ni-Mg-C compounds are covalent bonds structure.

The Ag 3d core level for Ag-$MgCNi_3$ is presented in Fig. 6. We saw that the core levels of Ag $3d_{3/2}$ and Ag $3d_{5/2}$ are at around 373.8 and 367.8 eV respectively, with a peak separation of 6.0 eV between those two peaks. Compared with the standard of silver [12], we found that the silver in the surface of Ag-$MgCNi_3$ is metal state except that there is a chemical shift of 0.5 eV for Ag-$MgCNi_3$. So we think that atomic Ag is in the vacancy between the lattices of $MgCNi_3$, and block the movement of electron-hole pairs.

Fig. 7 shows the resistance versus temperature curve of Ag-$MgCNi_3$. One can see that the superconductor transition temperature Tc is reduced at about 6.6 K and ΔTc about 0.6 K. The dR/dT equals $1.75 \times 10^{-3}$ Ω/K. Obviously, the superconductor transition temperature of Ag-$MgCNi_3$ is lower than that of $MgCNi_3$. The reason is that the atomic Ag diffuses to vacancy of $MgCNi_3$, forms some defects and lead to increase the density of electron state. According to BCS theory, the superconducting transition temperature depends on the density of electron-hole pairs. In our sample, Ag doped to the $MgCNi_3$, the results of the electron phonon coupling would be loss of the superconductivity.

4. Conclusions

In summary, the Ag-$MgCNi_3$ has been prepared by solid states reaction at 900 $^{o}$C for 2 hours. Their structure and surface features were studied by using XRD, Raman and XPS. It shows that the Ag-$MgCNi_3$ is $ABO_3$ type containing cubic Ag, and the Ag atomic is in the vacancy sites. The resisitivity measurement results show that the Ag diffusion to $MgCNi_3$ reduced superconductivity near 6.6 K. This is consistent with Cu doping in $MgCNi_3$.

**Acknowledgements:**

We would like to thank Prof. G. C. Che and Miss S. L. Jia for their assistance. This work reported here was supported by "Hundreds of Talents" program organized by the Chinese Academy of Sciences, P. R. China.

## Reference

[1] T. He, Q. Huang, A. P. Ramirez, Y. Wang, K. A. Regan, N. Rogado, N. P. Ong, R. J. Cava, Nature, 411(2001) 6833.

[2] C. Mazumdar, R. Nagarajan, C. Godart, L. C. Gupta, M. Latroche, S. K. Dhar, C. Levy Clement, B. D. Padalia, R. Vijayaraghavan, Solid State Commun. 87 (1993) 413.

[3] R. J. Cava, H. Takagi, H. W. Zanbergen, J. J. Krajewski, W. F. Peck Jr., T. Siegrist, B. Batlogg, R. B. Van Dover, R. J. Felder, K. Mizuhashi, J. O. Lee, H. Eisaki, S. Uchida, Nature, 367 (1994) 252.

[4] D. J. Singh, I. I. Mazin, cond-mat / 0105577, Phys. Rev. B, 64 (2001) R140507.

[5] H. Rosner, R. Weht, M. D. Johannes, W. E. Pickett, and E. Tosatti, Phys. Rev. Lett., 88 (2002) 027001.

[6] J. H. Shim, S. K. Kwon, and B. I. Min, Phys. Rev. B. 64(2001) 180510.

[7] Z. Q. Mao, M. M. Rosario, K. Nelson, K. Wu, I. G. Deac, P. Schiffer, Y. Liu, T. He, K. A. Regan, R. J. Cava, Cond-mat / 0105280.

[8] Q. Huang, T. He, K. A. Regan, N. Rogado, M. K. Haas, K. Inumaru, and R. J. Cava, Cond-mat / 0105240.

[9] M. A. Hayward, M. K. Haas, A. P. Ramirez, T. He, K. A. Regan, N. Rogado, K. Inumaru, R. J. Cava, Solid State Commun., 119 ( 2001 ) 491.

[10] A. Szajek, J. Phys. Condens. Matter. 13 (2001) L595.

[11] J. Q. Li, L. J. Wu, L. Li, Y. Zhu, Phys. Rev. B, 65 (2002) 052506.

[12] J. F. Moudler, W. F. Stickle, P. E. Sobol, K. D. Bomben, Handbook of X-ray Photoelectron Spectroscopy, Perkin-Elmer, Eden Praitie, MN, 1992.

Caption for Figures

Fig. 1 XRD pattern of Ag-$MgCNi_3$

Fig. 2 Raman scattering of Ag-$MgCNi_3$

Fig. 3 Mg 2p core level of Ag-$MgCNi_3$

Fig. 4 C1s core level of Ag-$MgCNi_3$

Fig. 5 Ni 2p core level of Ag-$MgCNi_3$

Fig. 6 Ag 3d core level of Ag-$MgCNi_3$

Fig. 7 The resistance versus temperature curve of Ag-$MgCNi_3$

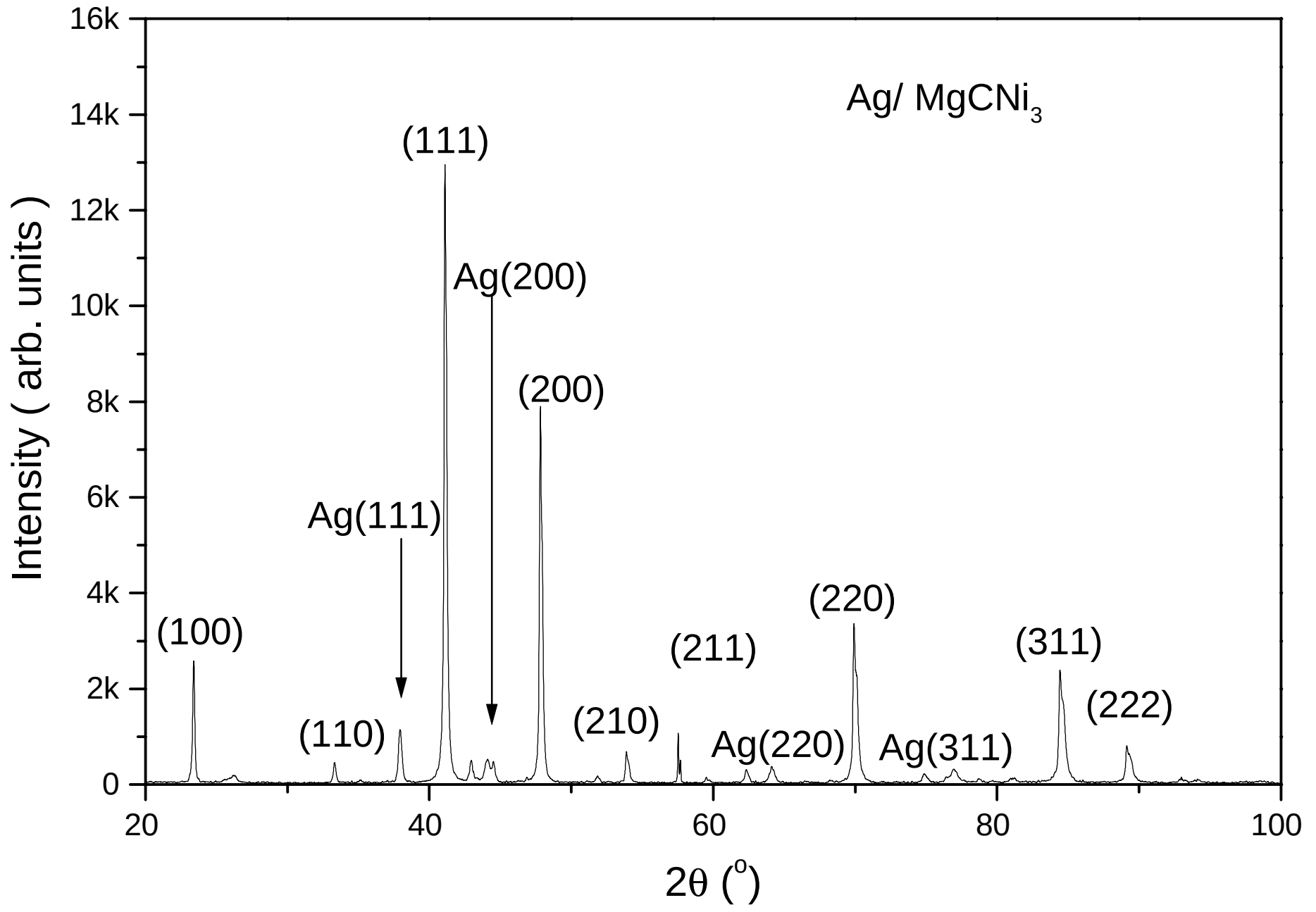

Fig. 1 XRD pattern of Ag-$MgCNi_3$

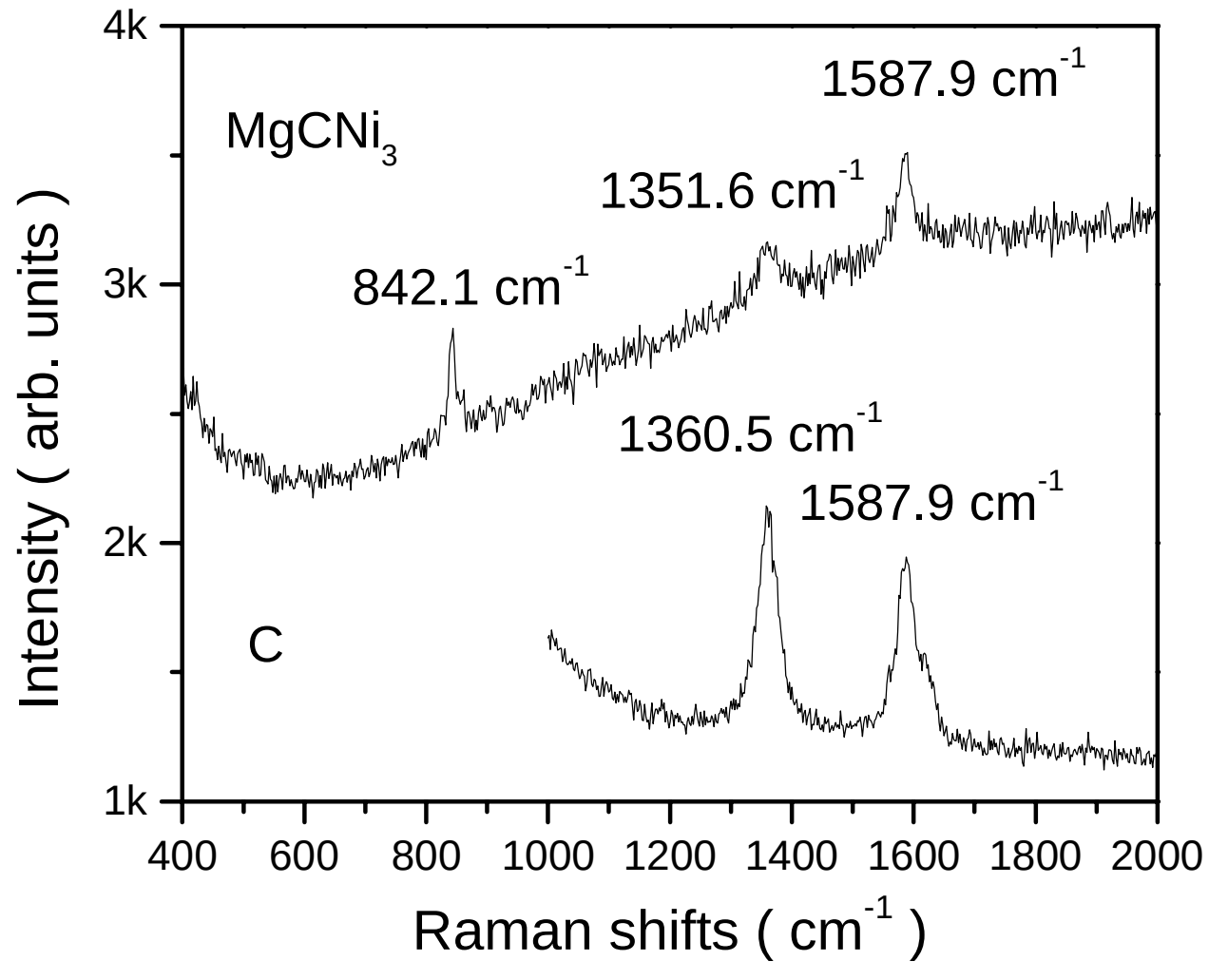

Fig. 2 Raman scattering of Ag-$MgCNi_3$

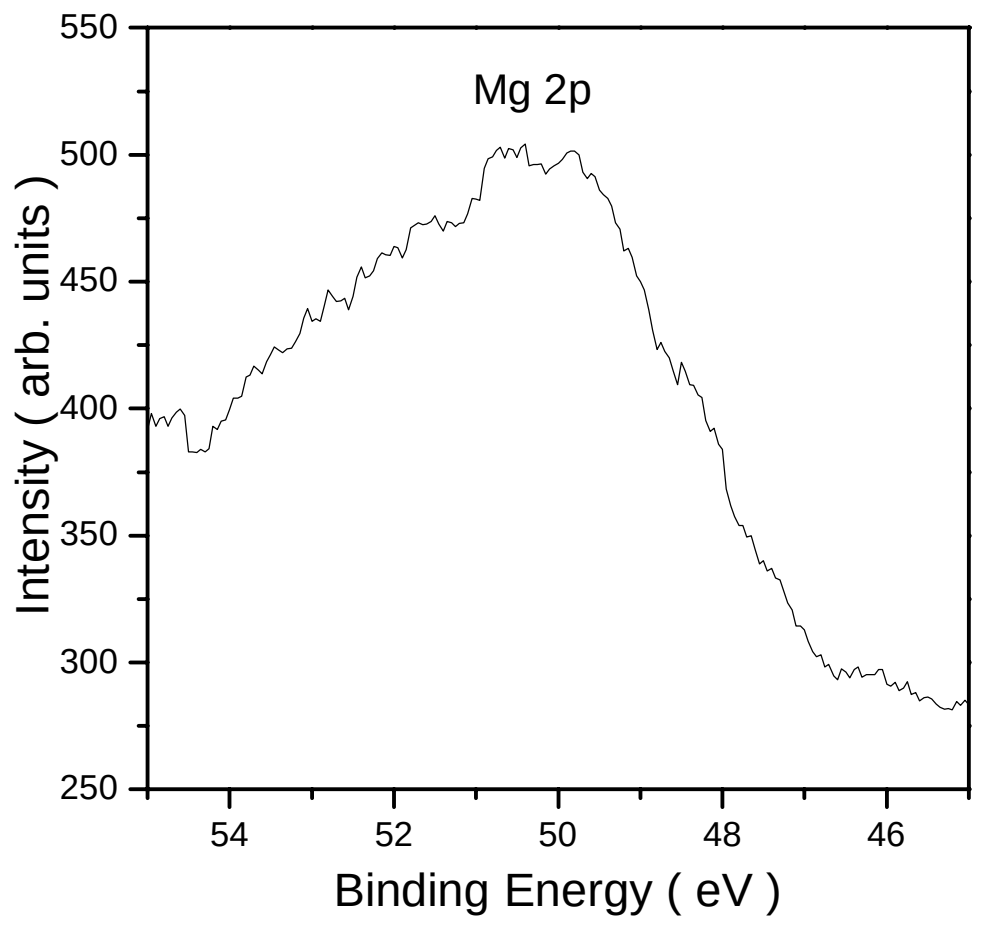

Fig. 3 Mg 2p core level of Ag-$MgCNi_3$

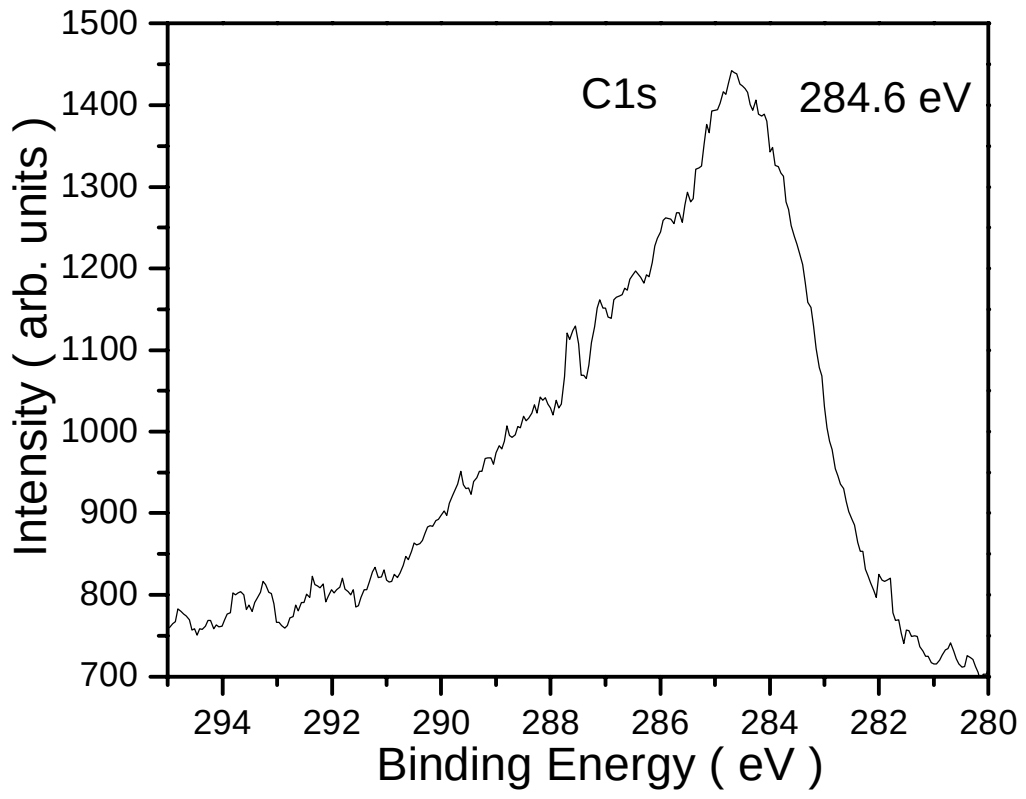

Fig. 4 C1s core level of Ag-$MgCNi_3$

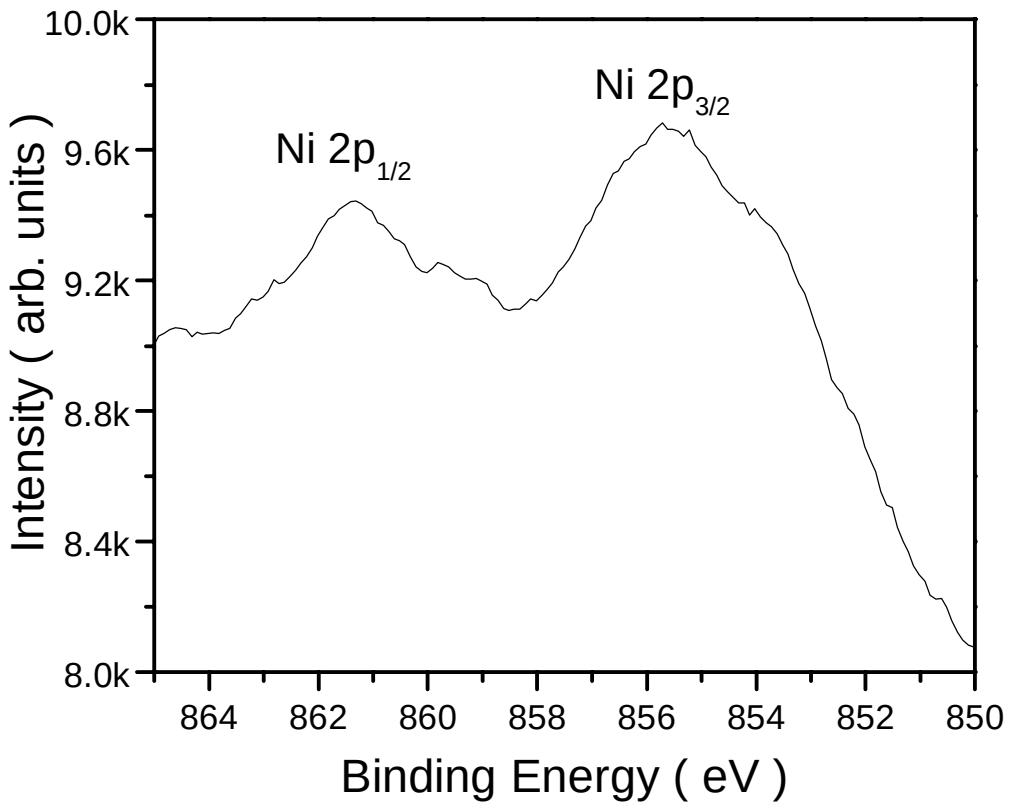

Fig. 5 Ni 2p core level of Ag-$MgCNi_3$

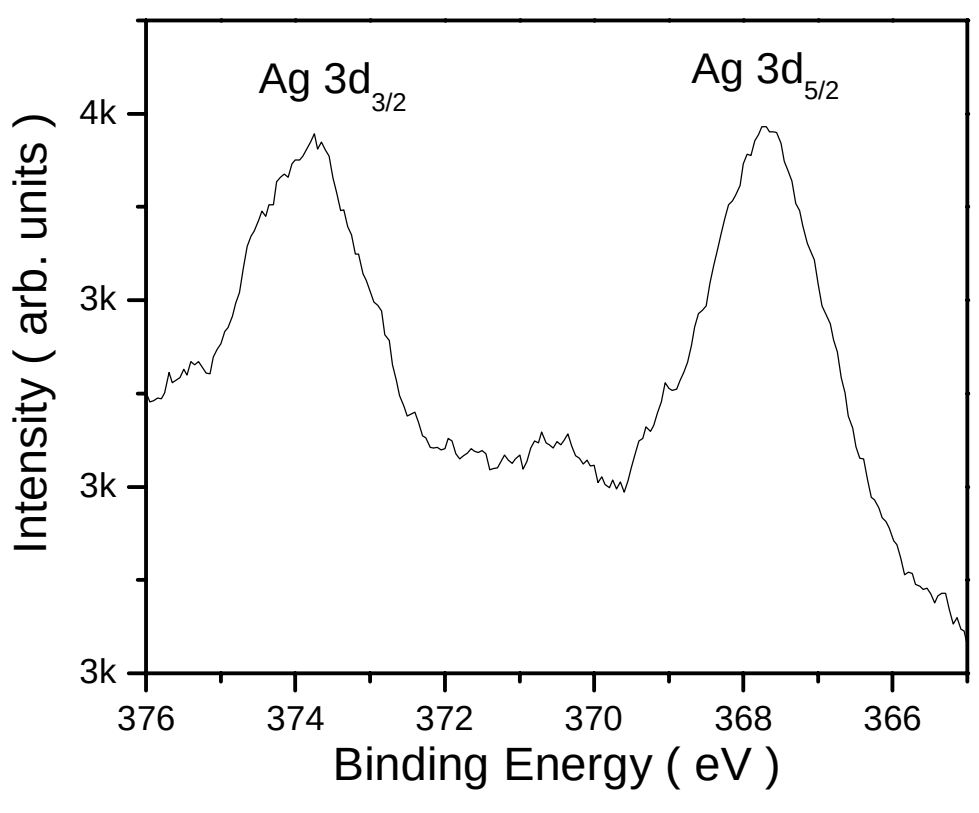

Fig. 6 Ag 3d core level of Ag-$MgCNi_3$

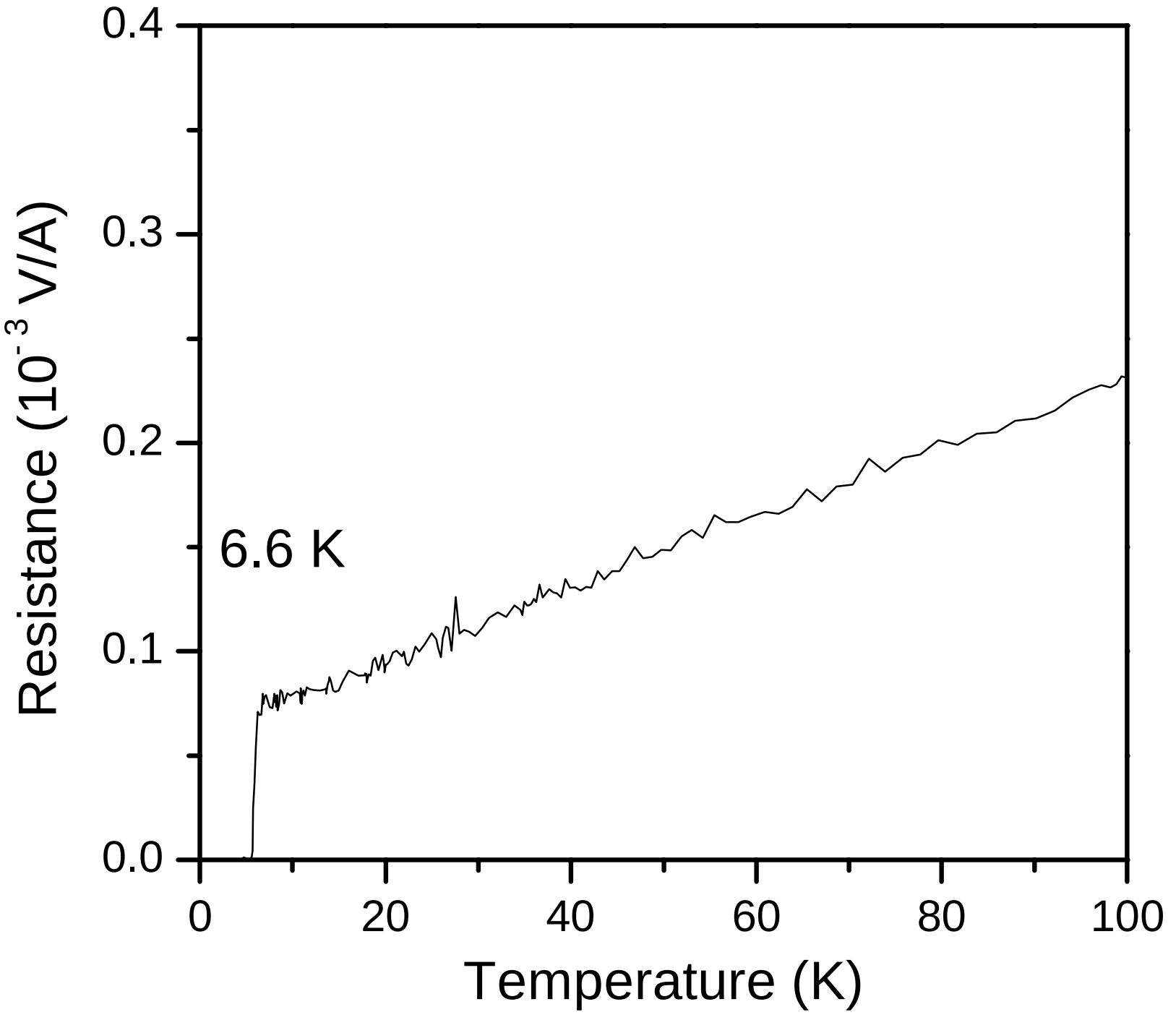

Fig. 7 The resistance versus temperature curve of Ag-$MgCNi_3$